\documentclass[prd,preprint,showpacs]{revtex4}
\usepackage{graphics}
\def \beq{\begin{equation}}
\def \eeq{\end{equation}}
\def \beqa{\begin{eqnarray}}
\def \eeqa{\end{eqnarray}}
\def \lms{\Lambda_{\overline{\scriptscriptstyle MS}}}

\def \ie{{\sl i.e.\/}}
\begin{document}

\title{Valence quarks in the QCD plasma: quark number susceptibilities and
   screening}
\author{Rajiv V.\ Gavai}
\email{gavai@tifr.res.in}
\author{Sourendu Gupta}
\email{sgupta@tifr.res.in}
\affiliation{Department of Theoretical Physics, Tata Institute of Fundamental
         Research,\\ Homi Bhabha Road, Mumbai 400005, India.}

\begin{abstract}
We have investigated the quark sector of quenched QCD for $1.5\le
T/T_c\le3$ in the continuum limit, using two different lattice
discretisations of quarks and extrapolating from lattice spacings between
$1/4T$ and $1/14T$.  At these temperatures, the flavour off-diagonal
susceptibility, $\chi_{ud}/T^2$, is compatible with zero at each lattice
spacing, and hence also in the continuum limit.  In the continuum limit,
the light quark susceptibilities are about 10\% less than the ideal gas
results even at the highest $T$, in agreement with hard thermal loop
predictions but marginally below a resummed perturbative computation.
For the mass range appropriate to the strange quark, the flavour
diagonal susceptibility is significantly smaller. Our estimate of the
Wroblewski parameter is compatible with observations at RHIC and SPS.
The continuum limit of screening masses in all (local) quark-bilinears
is very close to the ideal gas results.
\end{abstract}
\preprint{TIFR/TH/02-30, hep-lat/0211015}
\maketitle

\section{Introduction}

Heavy-ion experiments at RHIC are now analysing large quantities
of data and hope to identify and characterise the plasma phase of
QCD. As a result it has become imperative to put together an accurate
and detailed picture of this phase from theoretical analyses of QCD at
finite temperature. Present day lattice computations are able to do a
large part of this job. The reach of theory would certainly be enhanced if
perturbation theory is tested accurately against these computations and,
if found to work well, subsequently used in contexts where the lattice
is too ponderous to use.

The gluon sector of the QCD plasma has been extensively analysed.  In the
quenched theory the scaling of $T_c$ has been established to high accuracy
\cite{biel,myold}. The continuum limit of the equation of state has also
been determined to reasonably high accuracy \cite{biel,qcdpax}. This
shows strong departures from the ideal gas. It turns out that hard
thermal loop (HTL) resummation is not sufficient for a description
of the lattice results \cite{braaten}, nor is a Borel resummation of
the perturbative series \cite{parwani}. Other techniques such as an
approximately self-consistent resummation \cite{bir0} and dimensional
reduction \cite{helsinki} have been applied, and the former has been
successful in describing the lattice results for $T>2T_c$. In addition,
the spectrum of screening masses in the QCD plasma has been determined to
good accuracy \cite{saumen} and, for $T>2T_c$, found to be in agreement
with that determined non-perturbatively from a dimensionally reduced
model whose couplings are matched perturbatively to the four-dimensional
theory \cite{dimred}.  There is also direct evidence for the existence
of the gluon as a quasiparticle in the plasma for $T>1.5T_c$ \cite{gluon}.

The quark sector of the theory has not yet been explored in as much
detail. Part of the reason is that lattice computations with light
sea quarks are very expensive. While sea quarks are needed in the
neighbourhood of the phase transition, where they influence even the order
of the transition, sufficiently far away their effects are seen to be
small \cite{qnch,2fl}. In this paper we report on the physics of valence
quarks in the continuum limit of QCD for $T=1.5T_c$--$3T_c$, where sea
quark effects are marginal. The two aspects we concentrate on are quark
number susceptibilities \cite{gott} and screening correlators and masses.

Quark number susceptibilities \cite{gott,gavai,qnch,2fl,wrob,heller}
now occupy a position of high interest.  Due to their connection
with fluctuation phenomena \cite{koch} and strangeness production
\cite{wrob} they are useful inputs to relativistic heavy-ion collisions.
They are coefficients of the Taylor expansion of the free energy density
of QCD in terms of the chemical potential, and hence are useful checks on
recent attempts at numerical studies of finite-density QCD \cite{mu}. For
exactly the same reason they are a test-bed for suggested resummations
of the high-temperature perturbation series for QCD, which aim to
reproduce the free energy density, and hence the equation of state,
at high temperatures \cite{bir}. In this paper we report our study of
these susceptibilities at a large variety of lattice spacings and an
extraction of their continuum limit.

Screening correlators and masses are closely related to linear response
functions such as susceptibilities. They have been explored earlier at
finite lattice spacings, $1/4T$--$1/8T$. Staggered quarks were used
in 4 flavour \cite{4flav}, 2 flavour \cite{milc,2fl} and quenched
\cite{gocksch,plbold,boyd} simulations.  For reasons of symmetry a
large degeneracy of screening masses, of the zero temperature scalar
(S), pseudo-scalar (PS), and certain components of the vector (V)
and axial-vector (AV) quark bilinears, is expected in the QCD plasma
\cite{old}. This was not observed; instead lesser degeneracies were---
between parity partners S and PS or V and AV.  Similar results were also
obtained with overlap \cite{robert} and Wilson \cite{edwin} quarks in
quenched QCD, though the splitting between the S/PS and V/AV sectors
was seen to be substantially smaller. Here we study these correlators
on a wide range of lattice spacings and extrapolate the results to
the continuum.

In this paper we carefully address the question of the continuum limit
of these quantities in the quenched QCD plasma, where the valence quarks
are used as probes of the dynamics of gluons. To this end we perform
quenched computations over a range of temperatures and a large range of
lattice spacings, in order to perform a continuum extrapolation. We test
the extrapolation by working with two different formulations of quarks on
the lattice, with different systematics at finite lattice spacing. One
is the normal staggered formulation of quarks, the other a Symanzik
improved version of staggered quarks, called Naik quarks \cite{naik}. By
this means we identify a unique and verifiable continuum limit.

We would like to distinguish our approach from Symanzik (or loop level)
improvement of the action \cite{improve}. Such programs aim to approach
the continuum limit of the action on coarse lattices by adding terms to
the action which systematically, order by order in the lattice spacing
$a$, remove lattice artifacts up to some sufficiently high order in $a$.
When such a scheme is used in conjunction with improved operators,
it (ideally) obviates any need to go to very fine lattice spacings.
In this work we only use an improved operator for the valence quarks;
the action is not improved.  We reach the continuum limit by simulations
on a variety of lattice spacings and a smooth extrapolation.

In the next section we present definitions of all the quantities that
we measure, and results for these quantities for ideal quark gases
on a lattice. In Section 3 we present details of the simulations and
measurement schemes. The following section contains detailed results on
the lattice with discussion of the extrapolation to the continuum. The
continuum extrapolations are collected and discussed in the final section.

\section{Definitions}

The partition function of QCD with three quark flavours is
\beq
   Z(T,\mu_u,\mu_d,\mu_s) = \int{\cal D}U \exp[-S(T)]
       \prod_{f=u,d,s}\det M(T,m_f,\mu_f),
\label{part}\eeq
where the temperature $T$ determines the size of the Euclidean time
direction, $S(T)$ is the gluon part of the action, and the determinant
of the Dirac operator, $M$, contains as parameters the quark masses, $m_f$,
and the chemical potentials, $\mu_f$ for flavours $f$.  We also define
the chemical potentials $\mu_0=\mu_u+\mu_d+\mu_s$, $\mu_3=\mu_u-\mu_d$
and $\mu_8=\mu_u+\mu_d-2\mu_s$, which correspond to the diagonal flavour
$SU(3)$ generators. Note that $\mu_0$ is the usual baryon chemical
potential and $\mu_3$ is an isovector chemical potential.

The quark number density for flavour $f$, $n_f$, is defined as the
derivative of $F=\log Z/V$ ($Z$ is the partition function and $V$
the spatial volume) with respect to $\mu_f$.  The quark number
susceptibilities are the second derivatives
\beq
   \chi_{ff'}(T,\mu_u,\mu_d,\mu_s) = T
        \frac{\partial^2F}{\partial\mu_f\partial\mu_{f'}}.
\label{susc}\eeq
We will always drop the repeated subscript for the diagonal
susceptibilities. There are similar susceptibilities, $\chi_0$, $\chi_3$
and $\chi_8$ obtained by taking derivatives with respect to $\mu_0$,
$\mu_3$ and $\mu_8$ respectively.  Here we determine the susceptibilities
at $\mu_f=0$.  In this limit, all $n_f(T)=0$ but the susceptibilities
can be non-vanishing.

The flavour off-diagonal susceptibilities such as
\beq
   \chi_{ff'} = \frac TV \left\langle{\rm tr} M_f^{-1} M_f'\;
                     {\rm tr} M_{f'}^{-1} M_{f'}'\right\rangle
\label{offd}\eeq
are given solely in terms of the expectation values of quark-line
disconnected loops. Here $M_f$ denotes the Dirac operator for a
quark of flavour $f$, and $M_f'$ and $M_f''$ the first and second
derivatives with respect to $\mu_f$, taken term by term. The flavour
diagonal susceptibilities have contributions from both quark-line
connected and disconnected pieces---
\beq
   \chi_f = \frac TV \left[\left\langle\left({\rm tr} M_f^{-1} M_f'\right)^2
                     \right\rangle +
         \left\langle{\rm tr}
             \left(M_f^{-1} M_f'' - M_f^{-1}M_f'M_f^{-1}M_f'\right)
                  \right\rangle\right].
\label{chis}\eeq
We draw attention to the fact that the last two terms individually grow
rapidly with lattice volume, diverging in the infinite volume limit, such
that the difference is a finite quantity. This cancellation of divergences
is the result of a proper treatment of quark chemical potentials on the
lattice \cite{chempot}.  Numerically, the simplest quantity to evaluate
is the diagonal isovector susceptibility
\beq
   \chi_3 = \frac T{2V} \left\langle{\rm tr} \left(M_u^{-1} M_u'' 
                 - M_u^{-1}M_u'M_u^{-1}M_u'\right)\right\rangle,
\label{chi3}\eeq
where the factor half takes care of the isospin carried by each flavour.
In addition to this, we shall need the baryon number and electric charge
susceptibilities
\beq
   \chi_0 = \frac19\left(4\chi_3+\chi_s+4\chi_{ud}+4\chi_{us}\right)
       \qquad{\rm and}\qquad
   \chi_Q = \frac19\left(10\chi_3+\chi_s+\chi_{ud}-2\chi_{us}\right).
\label{chi0q}\eeq
Eqs.\ (\ref{offd}--\ref{chi3}) have been written per flavour of quarks, \ie,
the normalisation for the continuum ideal gas is $\chi_f/T^2=1$.

The screening correlators of local quark-bilinear (``meson'') operators
are defined as---
\beq
   C_\Gamma(z) = \sum_{x,y,t} \langle M^{-1}_{\alpha\beta}(x,y,z,t)\Gamma
        {M^\dag}^{-1}_{\beta\alpha}(x,y,z,t)\Gamma\rangle
\label{scr}\eeq
where $\Gamma$ denotes some Dirac-flavour structure, $\alpha$ and
$\beta$ are colour indices and $M^{-1}(r)$ is the inverse of the Dirac
operator for a point source in the fundamental of colour $SU(3)$ at the
origin. Since the partition function of eq.\ (\ref{susc}) can be expressed
as the trace of a spatial transfer matrix, the screening correlators
decay exponentially, with characteristic length scales called screening
lengths (inverse of screening masses, $\mu_\Gamma$). 

The finite temperature symmetries of such a spatial transfer matrix are
quite different from that of the Hamiltonian and have been worked out
in detail \cite{old}. In particular, it turns out that there are only
two independent local correlators. The combination of vector operators
${\rm V}_x-{\rm V}_y$ and axial vectors ${\rm AV}_x-{\rm AV}_y$ are in
the $B_1^{++}$ representation of the finite temperature symmetry group
$D_4^h$. \footnote{Here $V_\mu$ is the polarisation $\mu$ of the $T=0$
vector operator and ${\rm AV}_\mu$ is the polarisation $\mu$ of the $T=0$
axial vector.}.  All others, \ie, the scalar S, the pseudoscalar PS,
and all the components of the V and AV orthogonal to the $B_1^{++}$,
lie in the scalar representation, $A_1^{++}$.  This clearly implies
that the angular momentum $J$ is not a good quantum number for screening
correlators, and hence states of different $J$ can mix with each other.

The meson susceptibilities are the zero momentum part of the screening
correlator \cite{plbold} and can be written as
\beq
   \chi_\Gamma \equiv \langle{\rm tr} M^{-1}\Gamma M^{-1}\Gamma\rangle
   = \sum_i Z_i(\Gamma) \left(\mu_\Gamma^i\right)^{-2},
      \qquad{\rm where}\qquad
   Z_i = \left|\langle i|\overline\psi\Gamma\psi|0\rangle\right|^2,
\label{msus}\eeq
$|i\rangle$ denotes the eigenstate of the transfer matrix with
eigenvalue $\lambda_i$ and the screening masses are $\mu_\Gamma^i =
\log(\lambda_i/\lambda_0)$.  In the continuum limit, since $M'=\gamma_0$,
clearly $\chi_3 = \chi_{V_0}$; in fact, this identification underlies
the computation of \cite{bir}.  The PS
susceptibility is also involved in a chiral Ward identity---
\beq
   \langle\overline\psi\psi\rangle=m\chi_{PS}.
\label{ward}\eeq
At $T=0$, chiral symmetry is broken and the sum over states in
eq.\ (\ref{msus}) is dominated by the lowest term, due to the
vanishing Goldstone pion mass. Then this Ward identity shows that
$m_\pi^2\propto1/m$. However, for $T>T_c$, chiral symmetry is restored
and $\langle\overline\psi\psi\rangle$ should vanish in the chiral
limit. At the same time, $\chi_{PS}$ need not be dominated by one state
and $\mu_{PS}^i$ could depend weakly on $m$, going to a finite non-zero
limit as $m\to0$. It is interesting to note that since the screening
masses are equal for all the $A_1^{++}$ channels, the differences between
the susceptibilities can only come from the $Z_i$'s in eq.\ (\ref{msus}).

\begin{figure}[hbt]\begin{center}
   \includegraphics{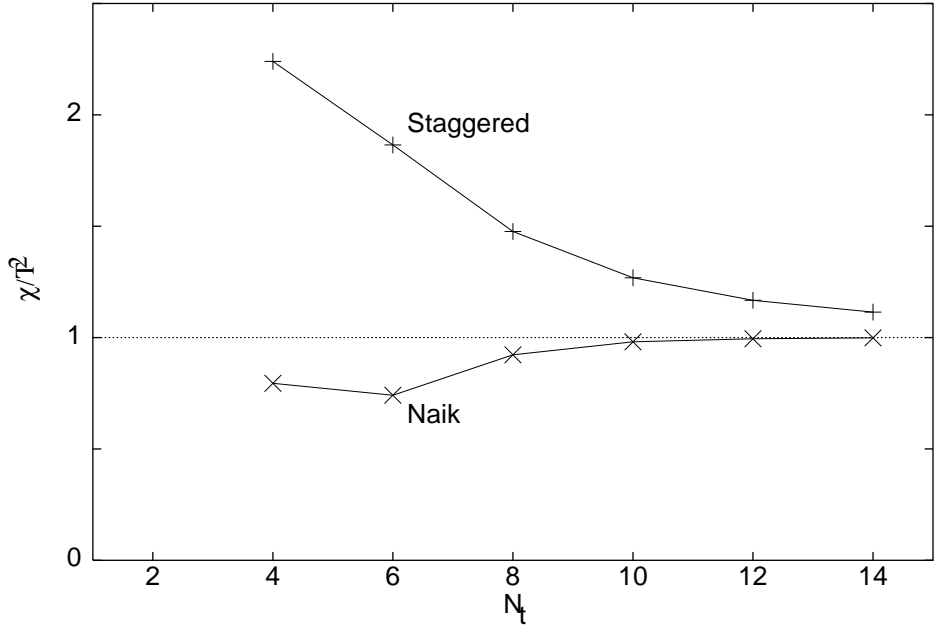}
   \end{center}
   \caption{The ideal gas on a lattice with $TV^{1/3}=3$ for
       staggered quarks (pluses) and Naik quarks (crosses), as a
       function of $N_t=1/aT$.}
\label{fg.ideal}\end{figure}

For an ideal gas of quarks on the lattice, the Dirac operator at zero chemical
potential is most simply written in momentum space where it is diagonal---
\beq
   M_{pq} = \delta_{pq}\left[ ma + i\gamma\cdot{\hat p} \right],
     \qquad{\rm where}\qquad
   {\hat p}_\mu = C_{10}\sin p_\mu + C_{30}\sin3p_\mu,
\label{ideal}\eeq
the momentum spectrum is $p_0=(2\pi/N_t)(n+1/2)$ with $1\le n\le N_t$ and
$p_i=2n\pi/N_i$ with $1\le n\le N_i$ in all other directions, and ${\hat
p}_\mu= \sin p_\mu$. For staggered quarks, where only a nearest neighbour
difference appears in the Dirac operator, $C_{10}=1/2$ and $C_{30}=0$. For
Naik quarks, on the other hand, $C_{10}=9/16$ and $C_{30}=-1/48$. The
construction of the number densities and conserved currents, and hence the
chemical potential on the lattice, is complicated for Symanzik improved
quarks, but goes through in the usual way \cite{naikpot}.

All the quantities of interest are exactly computable for an ideal quark
gas, not only in the continuum, but also on the lattice. In terms of
$1/D(p) = (ma)^2 + {\hat p}\cdot {\hat p}$, $p_\mu' = -(C_{10}\cos
p_0 + 3C_{30}\cos3p_0) \delta_{0\mu}$ and $p_\mu'' = (C_{10}\sin p_0
+ 9C_{30}\sin3p_0) \delta_{0\mu}$ we have---
\beqa
 \nonumber
   a^3\langle\overline\psi\psi\rangle_{FFT} &=& N_c\sum_p D(p),\\
 \nonumber
   a^2\chi_{FFT} &=& \frac{N_c}{N_tN_s^3}\sum_p \left\{
     D(p)\left[{\hat p}\cdot p'-p''\cdot p''\right] 
           + 2D^2(p)\left({\hat p}\cdot p''\right)^2\right\},\\
   a\mu_{FFT}^{stag} &=& 2 \sinh^{-1}\sqrt{ (ma)^2+\sin^2(\pi/N_t)}\;.
\label{stag}\eeqa
Figure \ref{fg.ideal} shows the susceptibilities for the staggered
and Naik quarks on a sequence of lattices with fixed aspect ratio
$N_s/N_t=3$. Note that the approach to the continuum limit is quite
different in the two cases. In free-field theory $\chi_{ud}=0$. The
computation of the correlators $C_\Gamma(z)$ in the ideal gas is also
straightforward. Perturbative expansions in the gauge coupling, $g$,
have been performed in the continuum. $\chi_3$ has been computed in
the hard thermal loop (HTL) approximation to order $g^3$ as well as in
a skeleton graph resummation \cite{bir}. To the best of our knowledge,
there has been no such computation of screening correlators, screening
masses or the remaining susceptibilities.

\begin{table}\begin{center}\begin{tabular}{r|lll|lll|lll}
   \hline
   $N_t$ & \multicolumn{3}{|c}{$1.5T_c$} 
         & \multicolumn{3}{|c}{$2T_c$} 
         & \multicolumn{3}{|c}{$3T_c$} \\
   \cline{2-10}
         & $\beta$ & $V$ & ${\rm Stats}$ &
           $\beta$ & $V$ & ${\rm Stats}$ &
           $\beta$ & $V$ & ${\rm Stats}$ \\
   \hline
     4 & 5.8941 & $12^3$ & $50\times{\bf 51}$ &
         6.0625 & $12^3$ & $50\times{\bf 50}$ &
         6.3384 & $16^3$ & $50\times{\bf 28}$ \\
     6 & 6.0625 & $20^3$ & $1000\times{\bf 54}$ &
         6.3384 & $20^3$ & $1000\times{\bf 55}$ &
         6.7    & $20^3$ & $1000\times{\bf 59}$ \\
     8 & 6.3384 & $18^3$ & $1000\times{\bf 57}$ &
         6.55   & $18^3$ & $100\times{\bf 54}$ &
         6.95   & $26^3$ & $500\times{\bf 30}$ \\
       &        & $32\times18^2$ & $100\times{\bf 47}$ &
                & $32\times18^2$ & $100\times{\bf 20}$ &
                & $32\times26^2$ & $100\times{\bf 20}$ \\
    10 & 6.525  & $22^3$ & $50\times{\bf 105}$ &
         6.75   & $22^3$ & $50\times{\bf 219}$ &
         7.05   & $32^3$ & $50\times{\bf 23}$ \\
       &        & $40\times22^2$ & $100\times{\bf 20}$ &
                & $40\times22^2$ & $100\times{\bf 20}$ &
                & $40\times32^2$ & $100\times{\bf 20}$ \\
    12 & 6.65   & $26^3$ & $50\times{\bf 75}$ &
         6.90   & $26^3$ & $50\times{\bf 173}$ &
         7.20   & $38^3$ & $50\times{\bf 50}$ \\
       &        & $48\times26^2$ & $100\times{\bf 24}$ &
                & $48\times26^2$ & $100\times{\bf 61}$ &
                & $48\times38^2$ & $100\times{\bf 20}$ \\
    14 & & & & 7.00 & $30^3$ & $1000\times{\bf 48}$ & & & \\
   \hline
\end{tabular}\end{center}
\caption{The lattice sizes, $N_t\times N_s^3$, Wilson coupling, $\beta$, and
   statistics used in this study. The statistics is reported as $N_{sep}
   \times {\mathbf N_{stat}}$ where $N_{sep}$ is the number of sweeps between
   measurements and ${\mathbf N_{stat}}$ is the number of measurements. 1000
   initial sweeps were discarded for thermalisation in all cases except the
   $N_t=14$ run, where 7000 sweeps were discarded. For asymmetric spatial
   sizes the long direction is called the $z$-direction.}
\label{tb.runs}\end{table}

\section{Methods}

As already mentioned, we have investigated quark number susceptibilities
using two different realisations of lattice quarks.  For staggered quarks,
the derivative in the Dirac operator is discretised using a one-link
separated finite difference operator. This has a discretisation error
of order $a^2$. The improvement suggested by Naik \cite{naik} is to
take a specific three-link term (see eq.\ \ref{ideal}) which cancels
the leading term of the discretisation error for staggered Fermions
up to ${\cal O}(a^3)$.  This possible improvement comes at a fourfold
increase in computational cost, due to the more complicated structure
of the Dirac operator.

Since the discretisation error in the evaluation of the susceptibilities
comes partly from the gauge action and partly from the Dirac operator,
and we use the Wilson gauge action, the part of the ${\cal O}(a^2)$
error coming from this source would remain the same in the two
determinations. However, if a large part of the ${\cal O}(a^2)$ error
in the determination of susceptibilities using staggered quarks comes
from the traces of the Dirac operator in eqs.\ (\ref{offd}--\ref{chi3}),
then the slope in the extrapolation to the continuum limit should be
substantially different between the two quark formulations. That this is
so can be seen already in the computation for the ideal quark gas,
illustrated in Figure \ref{fg.ideal}.

In this paper we report results on susceptibilities and screening
measured on lattices with $4\le N_t\le14$ for $T=1.5T_c$, $2T_c$
and $3T_c$.  The lattices sizes, couplings and statistics used are
reported in Table \ref{tb.runs}. Due to statistical errors in the
determination of critical couplings various other quantities that go
into a scale determination, the temperature corresponding to a given
lattice coupling may be in error by about 5\% \cite{old}. The simulations
have been performed with a Cabbibo-Marinari pseudo-heatbath technique
with 3 $SU(2)$ subgroups updated on each hit, each sweep consisting
actually of 5 sweeps of this algorithm.  For susceptibilities we have
made measurements using both staggered and Naik quarks at $m/T_c=0.1$,
$0.5$ and 1. For staggered quarks we have also used $m/T_c=0.03$. For
screening masses and correlators we have used only staggered quarks.

\begin{figure}[hbt]\begin{center}
   \includegraphics{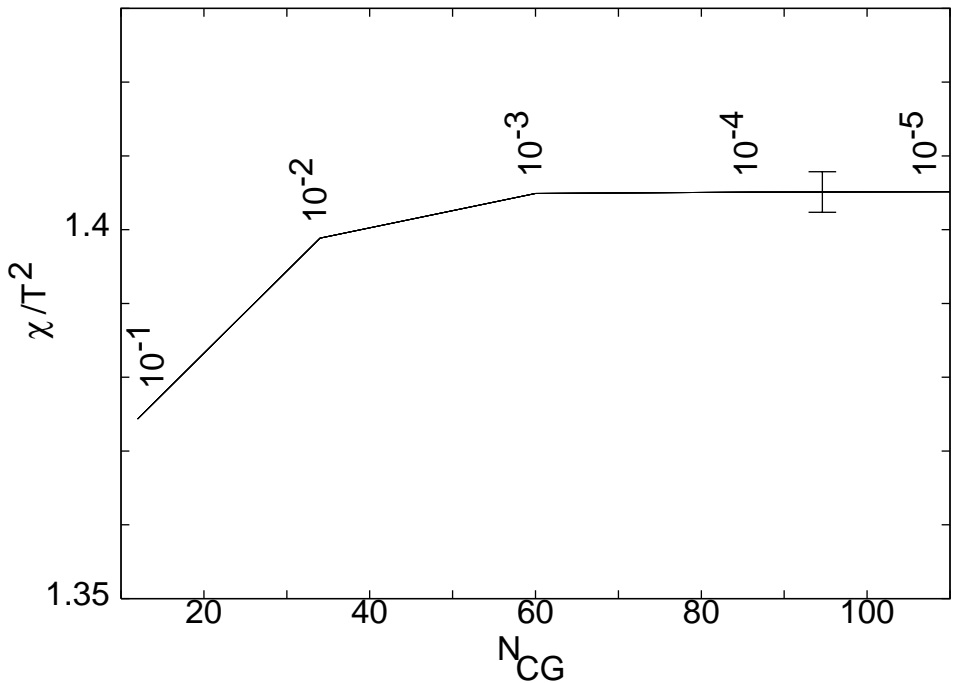}
   \end{center}
   \caption{The variation of the estimate of $\chi_3/T^2$ on a single
       configuration with $N_{CG}$, which values as $\ln(1/\epsilon_{CG})$.
       The values of $\epsilon_{CG}$ are marked at appropriate points
       on the curve. The size of typical errors on the estimates, after
       averaging over 50 gauge configurations with $N_v=10$, is shown as
       the bar. This is insensitive to $\epsilon_{CG}$.}
\label{fg.prec}\end{figure}

The stochastic evaluation of traces depends on the identity
\beq
   {\cal I}_{\alpha\beta} = \frac1{2N_v}\sum_{i=1}^{N_v} R_i^\alpha (R_i^*)^\beta,
\label{identity}\eeq
where $\cal I$ is the identity matrix, $R_i^\alpha$ is the component $\alpha$
of the vector $R_i$, which is one of a set of $N_v$ uncorrelated vectors with
complex components drawn independently from a Gaussian ensemble with unit
variance, and the star denotes complex conjugation. This leads to the stochastic
estimators,
\beq
   {\rm Tr}\,A = {\rm Tr}\,A{\cal I}= \frac1{2N_v}\sum_{i=i}^{N_v} R_i^\dag AR_i.
   \qquad\qquad ({\rm Tr}\,A)^2 = {\rm Tr} A{\cal I}A{\cal J}
\label{stoch}\eeq
where the vectors used in the stochastic estimators of the two identity
matrices, $\cal I$ and $\cal J$, are different. This is done by dividing
the set of $R_i$ into disjoint blocks.

It turns out that the divergence cancellation in eq.\ (\ref{chi3}) is
numerically the hardest part of the computation. In previous works we
had taken a half-lattice version of the staggered Dirac operator, which
performed extremely well on small lattice volumes but was inefficient at
divergence cancellation on large volumes \cite{wrob}, leading to enormous
statistical fluctuations. We found that the full lattice version of the
staggered Dirac operator performs much better in this respect, and has
no trouble handling our largest lattices.  The statistical fluctuations
are under good control already with $N_v=10$, and one has no need to look
for reduced variance versions of eq.\ (\ref{identity}). We found that it
is important to work in that sector of the quenched theory in which the
Wilson-Polyakov line, $L$, is purely real. This can always be achieved
by a $Z_3$ rotation in the quenched theory.

The slowest part of the computation is the inversion of the Dirac
operator, $M$, which is done by a conjugate gradient iteration. The
time taken is linear in the number of iterations, $N_{CG}$. This in turn
depends on the precision required in the solution, which is specified by
a single number $\epsilon_{CG}$ through the requirement that $|Mx-r|^2\le
N_tV\epsilon_{CG}$ (here $r$ is one of the random vectors and $x$ is the
estimate of $M^{-1}r$). In Figure \ref{fg.prec} we show how the estimate
of $\chi_3/T^2$ changes with $N_{CG} \propto \log (1/\epsilon_{CG})$. It
is clear that with $\epsilon_{CG}=10^{-3}$ the result is statistically
indistinguishable from that with $\epsilon_{CG}=10^{-5}$, but the solution
is achieved at half the CPU cost. We have used $\epsilon_{CG}=10^{-3}$
in all our computations of $\chi$. It turns out that the estimation
of screening correlators is more sensitive to the stopping criterion,
requiring $\epsilon_{CG}=10^{-5}$.

The fits and fit errors involved in the continuum extrapolations are
done by a bootstrap method. In this, bootstrap samples of the data are
extracted and fits are performed on statistics obtained from each such
sample.  The statistics of the fitted parameters are then used to obtain
non-parametric estimators of the mean and variance.  We have varied the
number of bootstrap samples by two orders of magnitude, between 10 and
1000, to check that the estimators are stable.

We have used staggered quarks to determine screening correlators and
masses. Recall the well-known fact that for these, the one link transfer
matrix contains complex eigenvalues and leads to oscillations in the
screening correlators.  However, the two link transfer matrices have
real eigenvalues, and the corresponding states can be projected by a
simple parity projection.  Due to the three-link terms in the Naik Dirac
operator, real eigenvalues are obtained only for the four-link transfer
matrix.  The one link transfer matrix contains oscillatory mixtures of
various states from which pure states need to be projected out. We show
in the next section that we can reach the continuum limit with staggered
quarks. Since they lead to simpler correlators, we have analysed screening
phenomena only with these. We have concentrated on degeneracies of the
screening correlators as well as the screening masses. The latter were
extracted from local masses as well as by making two-mass fits to the
parity projected correlators.  Details of these techniques remain the
same as in \cite{2fl}.  For bias-free extraction of screening masses,
large separations in the spatial directions are necessary. Consequently,
we have only used data from lattices with $N_z=4N_t$ for this analysis
(see Table \ref{tb.runs}).

\section{Lattice Results}

\begin{table}[htb]\begin{center}\begin{tabular}{r|ll|l|ll|l|ll|l}
   \hline
   $N_t$ & \multicolumn{3}{|c}{$1.5T_c$} 
         & \multicolumn{3}{|c}{$2T_c$} 
         & \multicolumn{3}{|c}{$3T_c$} \\
   \cline{2-10}
         & \multicolumn{2}{|c}{Staggered} & Naik 
         & \multicolumn{2}{|c}{Staggered} & Naik 
         & \multicolumn{2}{|c}{Staggered} & Naik \\
   \cline{2-10}
         & $0.03$ & $0.1$ & $0.1$ & $0.03$ & $0.1$ & $0.1$ & $0.03$ & $0.1$ & $0.1$  \\
   \hline
    4 & 1.816 (16) && 1.112 (22) 
      & 1.846  (6) && 1.137 (14) 
      & 1.911  (5) && 1.114 (15) \\
$6^*$ & 1.312 (59) && 0.957 (36) 
      & 1.428  (5) & 1.425 (5) & 1.033 (14) 
      & 1.492  (4) && 0.993 (10) \\
    8 & 1.110 (11) & 1.105 (11) & 0.936 (9)
      & 1.144 (11) & 1.142 (11) & 0.952 (8) 
      & 1.191 (19) & 1.190 (19) & 0.979 (5) \\
   10 & 1.014  (6) & 1.009  (6) & 0.921 (7) 
      & 1.060  (8) & 1.058  (8) & 0.948 (4) 
      & 1.066  (7) & 1.066  (7) & 0.956 (7) \\
   12 & 0.973 (14) & 0.969 (14) & 0.915 (7) 
      & 0.994  (7) & 0.992  (7) & 0.941 (9) 
      & 1.027  (6) & 1.026  (6) & 0.950 (5) \\
   14 &&&            
      & 0.985 (10) && 0.949 (2)
      &&&\\
   \hline
\end{tabular}\end{center}
\caption{Lattice results for $\chi_3/T^2$, with $m/T_c=0.03$ and $0.1$ for staggered quarks
   and $0.1$ for Naik quarks. For $N_t=6$ the three temperatures are $1.33T_c$, $2T_c$ and
   $3.13T_c$.}
\label{tb.lat}\end{table}

\begin{figure}[t!]\begin{center}
   \scalebox{0.6}{\includegraphics{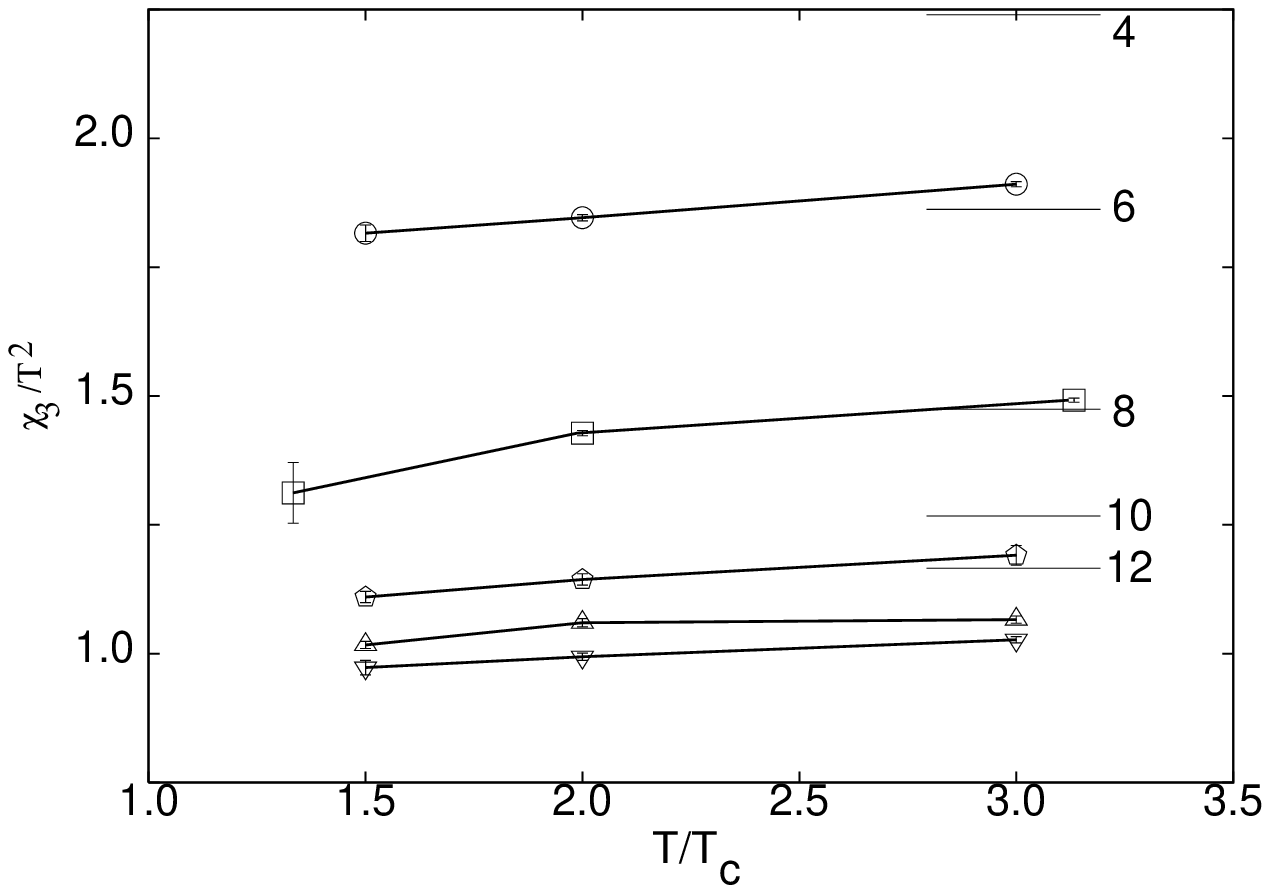}}
   \scalebox{0.6}{\includegraphics{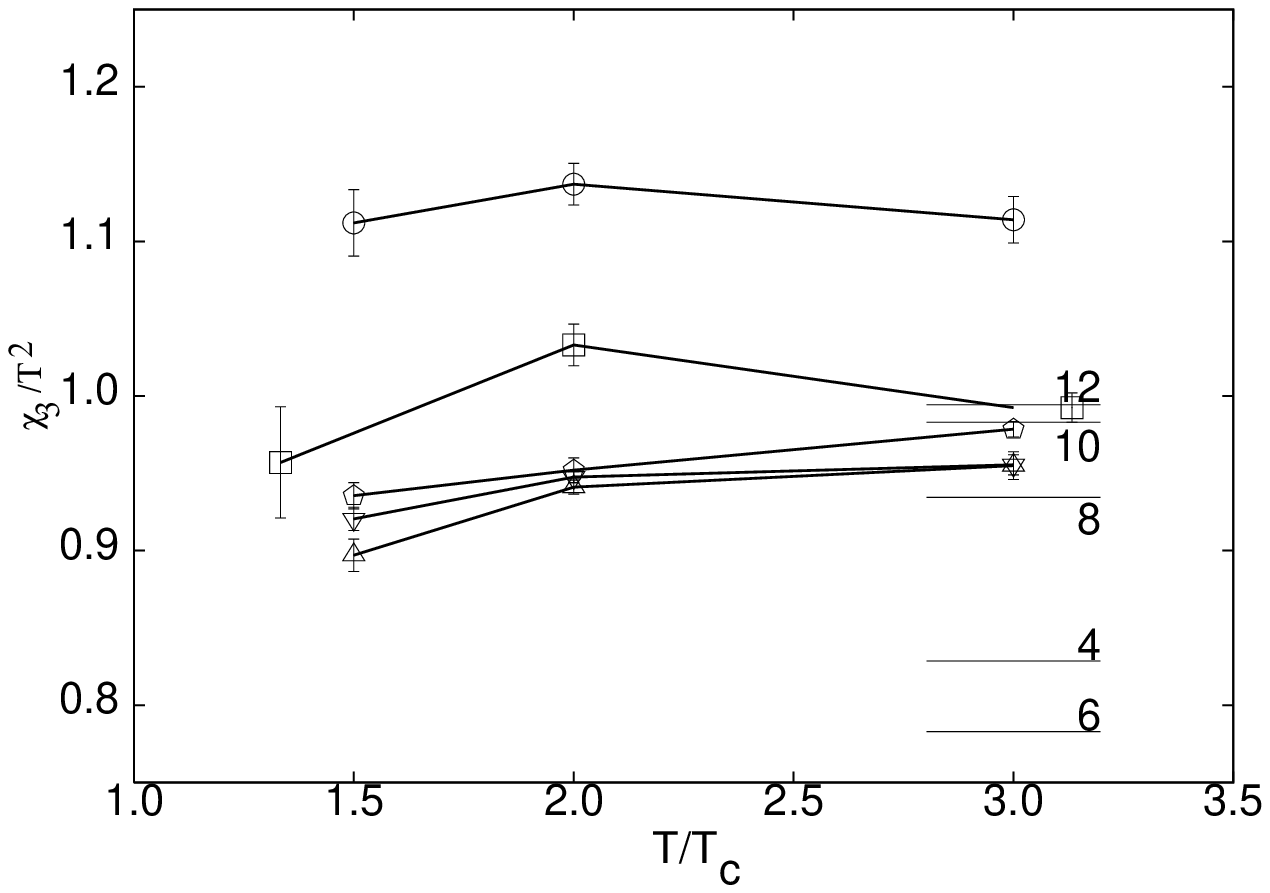}}
   \end{center}
   \caption{The temperature dependence of $\chi_3/T^2$ with staggered
      (left panel) and Naik (right panel) quarks of masses $0.03T_c$
      and $0.1T_c$ respectively, with $N_t=4$ (circles),
      6 (boxes), 8 (pentagons), 10 (down triangles) and 12 (up
      triangles).  Note the differences in the y-axes in the two cases.
      The bold lines join the centers of the data points. The thin lines
      show the value of $\chi_3/T^2$ for an ideal gas on the same
      lattices at $3T_c$.}
\label{fg.naiktdep}\end{figure}

As shown in Figure \ref{fg.ideal} an ideal gas of staggered quarks has
strong lattice artifacts and one needs lattices with $N_t\simeq8$--12
in order to get continuum results. Computations in quenched QCD, shown
in Figure \ref{fg.naiktdep} also indicate this. Note specially that for
$N_t=4$ and 6, $\chi_3/\chi_{FFT}$ decreases with $N_t$ at fixed $T$,
unlike at larger $N_t$ where it increases. This means that with staggered
quarks one needs fairly fine lattices for a smooth extrapolation to
the continuum.

\begin{figure}[b!]\begin{center}
   \includegraphics{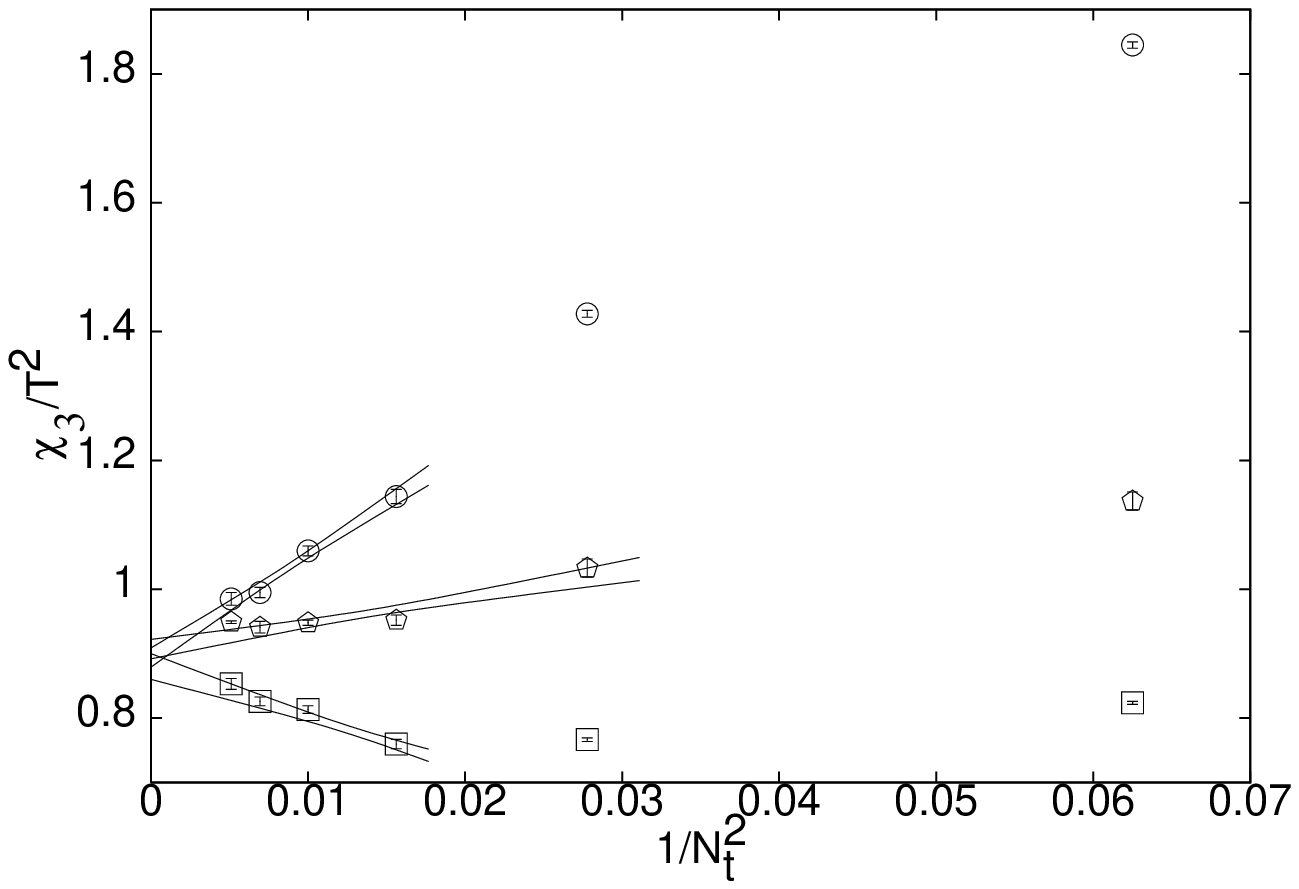}
   \end{center}
   \caption{The continuum extrapolation of staggered and Naik quark
       data for $\chi_3$ at $2T_c$. The circles are the extrapolation
       of $\chi_3/T^2$ for staggered quarks, the boxes for
       $\chi_3/\chi_{FFT}$ for staggered quarks, and the pentagons
       for $\chi_3/T^2$ for Naik quarks. The bands are 1--$\sigma$
       error bands for an extrapolation in $a^2\propto1/N_t^2$.}
\label{fg.extrap}\end{figure}

Unfortunately, Naik quarks also have their own quirks. An ideal
gas of Naik quarks on a lattice approaches the continuum smoothly
only for $N_t>6$ as shown in Figure \ref{fg.ideal}.  The interacting
theory also exhibits departures from smoothness, as shown in Figure
\ref{fg.naiktdep}. At $1.5T_c$ it might seem that the $N_t=6$ and 8
lattices scale, but at $2T_c$ this scaling is seen to be accidental. In
turn, at $2T_c$ one sees $N_t=8$, 10 and 12 as scaling, but at $3T_c$
this turns out to be accidental.

\begin{figure}[hbtp]\begin{center}
   \scalebox{0.8}{\includegraphics{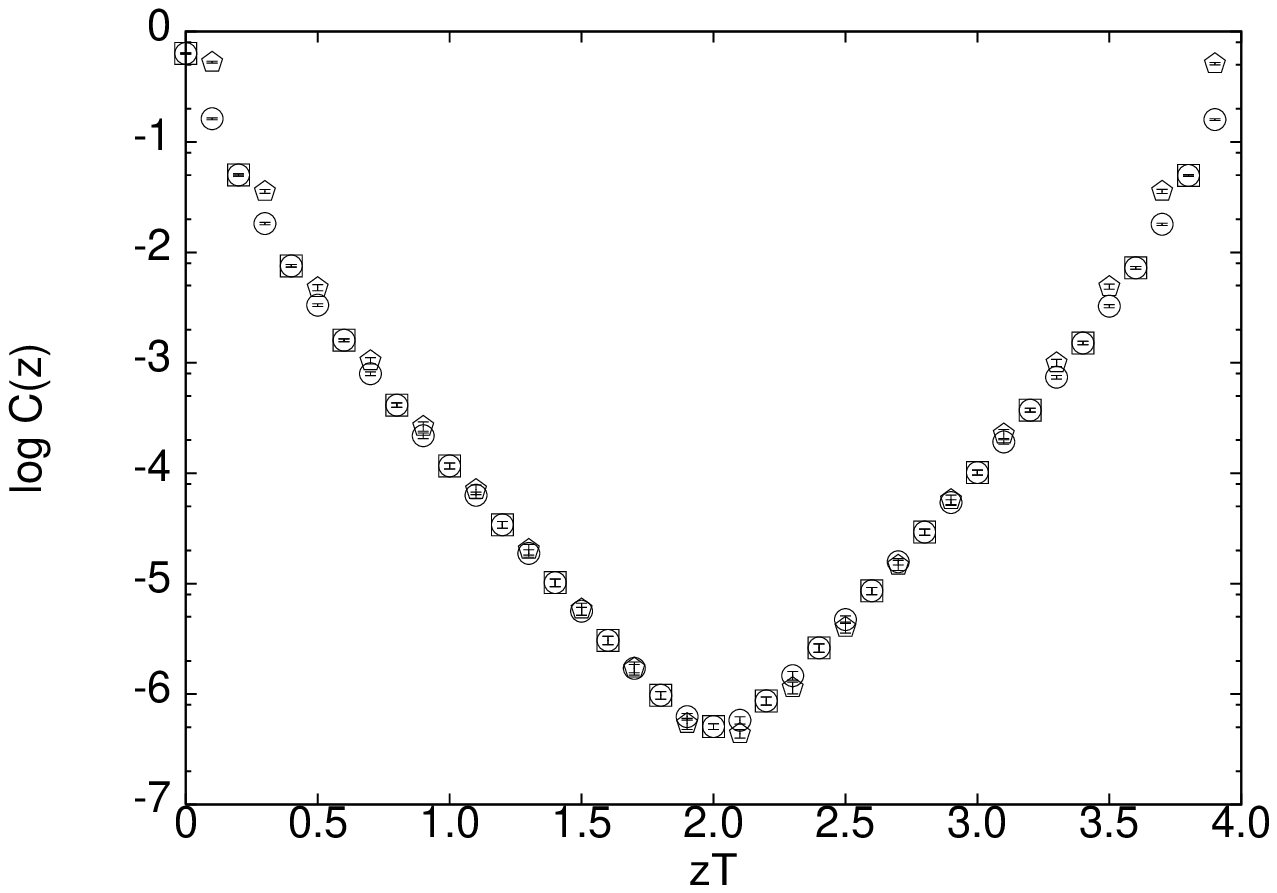}}
   \end{center}
   \caption{The parity projected screening correlators evaluated
      with staggered quarks of mass $0.03T_c$ at $T=2T_c$. The degeneracy
      of PS (circles) and S (boxes) correlators is evident, as is the
      equality of the screening masses for these and the V (pentagons)
      correlator. The S and PS correlators are divided by 3 to
      bring them into coincidence with the V. The other $A_1^{++}$ correlators,
      which are not shown, are also in equally good agreement with these.}
\label{fg.coincide}\end{figure}

\begin{figure}[htbp]\begin{center}
   \scalebox{0.8}{\includegraphics{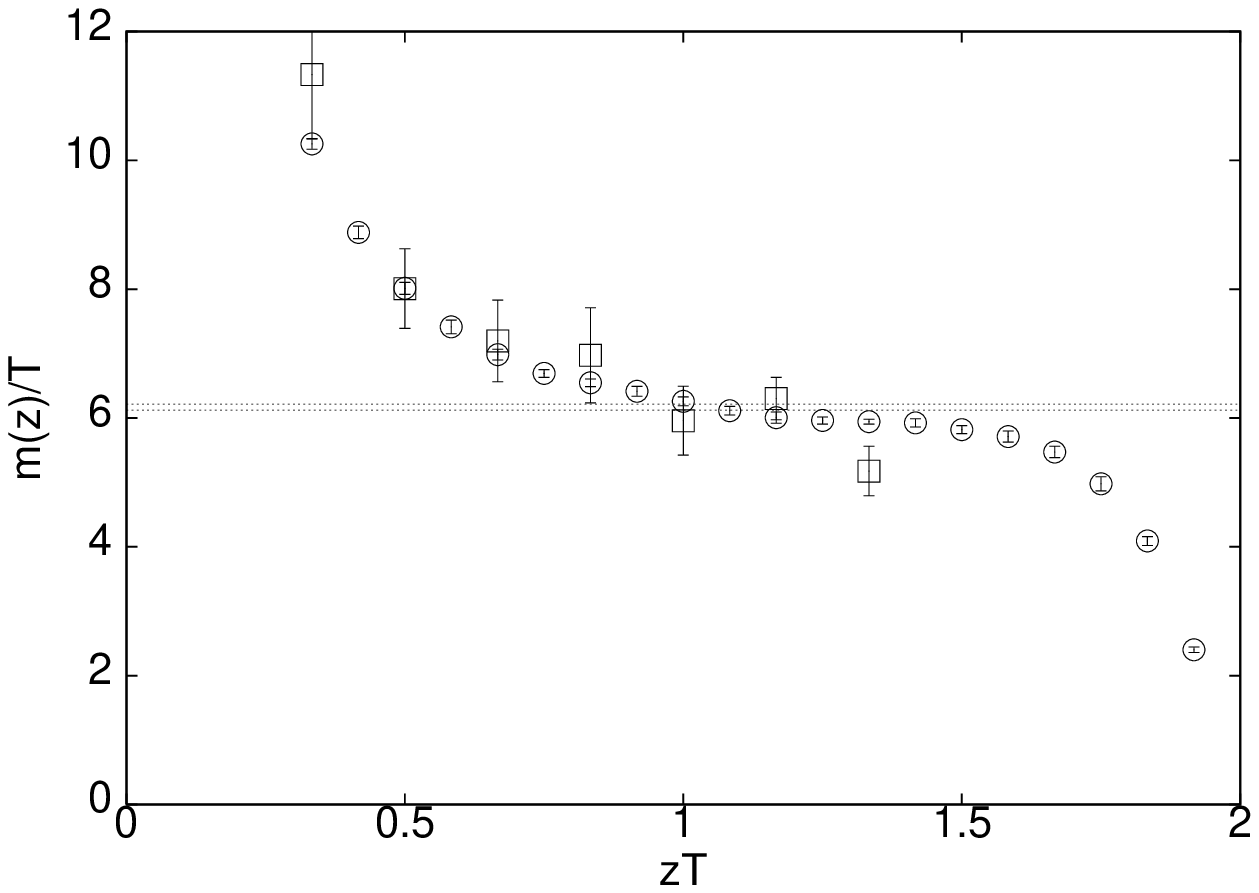}}
   \end{center}
   \caption{Local masses for the PS correlator (circles) compared with the
      result of a two-mass fit. The data are for a $12\times48\times26^2$
      lattice at $2T_c$ with $m=0.03T_c$ for staggered quarks. Also shown
      are local masses for the V correlator (boxes) in the $A_1^{++}$
      representation.}
\label{fg.mpi}\end{figure}

Our results for $\chi_3/T^2$ obtained with staggered and Naik quarks for
the lightest quark masses are collected in Table \ref{tb.lat}.  In Figure
\ref{fg.extrap} we show the extrapolation of staggered and Naik quark
results on $\chi_3$ at $T=2T_c$ to the continuum limit. For staggered
quarks only the data for $N_t\ge8$ fits a simple quadratic extrapolation
of the form $\chi_3(a)/T^2=\chi_3/T^2+s/N_t^2$ \footnote{Equally good
fits are obtained to the form $\chi_3(a)/T^2=\chi_3/T^2(1+s'/N_t^2)^{-1}$.}.
In fact, for smaller $N_t$ the extrapolations of $\chi_3/\chi_{FFT}$ even
gives the wrong sign for $s$ \footnote{Since the study in \cite{wrob}
used only $N_t\le8$, the extrapolation of $\chi_3/T^2$ led to strong
underestimation of the continuum limit. Since the extrapolation
affected $\chi_3$ and $\chi_s$ in the same way, the results for the
Wroblewski parameter, $\lambda_s$, were almost unaffected, as this work
verifies. The results for $\chi_{ud}/T^2$ were unrelated, and hence
unaffected.}.  However, for $N_t\ge8$ the extrapolations of $\chi_3/T^2$
and $\chi_3/\chi_{FFT}$ give identical results.  Since the two ratios
reach this limit from different directions, their agreement is already a
good indicator of having the right continuum limit.  For Naik quarks at
$2T_c$, the quadratic extrapolation looks reasonable for $N_t\ge4$. This
is accidental; at $3T_c$ or at $1.5T_c$ a good fit is obtained only
for $N_t\ge6$. Furthermore, as already mentioned, the extrapolation
of $\chi_3/\chi_{FFT}$ for Naik quarks does not coincide with that
of $\chi_3/T^2$ even at the 3-$\sigma$ level. The extrapolation of
$\chi_3/T^2$ does, however, agree within errors with the continuum
extrapolation from staggered quarks.

\begin{table}[htb]\begin{center}\begin{tabular}{r|lll}
   \hline
   $N_t$ & $\qquad1.5T_c$ & $\qquad2T_c$ & $\qquad3T_c$ \\
   \hline
   4 & $\qquad$4.12 (12)  & $\qquad$4.576 (16) & \\
   6 &            & $\qquad$5.34 (6)   & \\
   8 & $\qquad$5.536 (24) & $\qquad$5.744 (32) & $\qquad$5.928 (64) \\
  10 & $\qquad$5.67 (2) & $\qquad$6.07 (4) & $\qquad$6.04 (10) \\
  12 & $\qquad$5.904 (96) & $\qquad$6.156 (48) & $\qquad$6.156 (180) \\
   \hline
\end{tabular}\end{center}
\caption{$A_1^{++}$ local masses for staggered quarks with $m=0.03T_c$.}
\label{tb.mass}\end{table}

\begin{figure}[htbp]\begin{center}
   \includegraphics{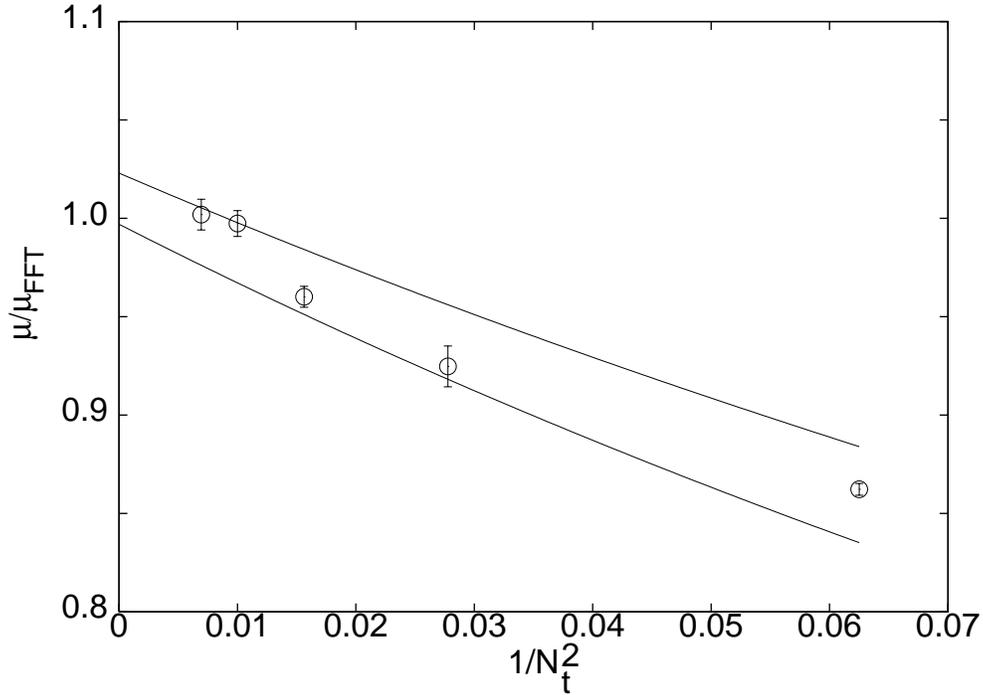}
   \end{center}
   \caption{Extrapolating the ratio of the measured $A_1^{++}$ screening
      mass, $\mu_{PS}/T$, and its value in an ideal quark gas to the
      continuum limit. The data are for $T=2T_c$ and $m=0.03T_c$ for
      staggered quarks. Already at lattice spacings $a=1/10T$ and greater
      the result is compatible with unity.}
\label{fg.mass}\end{figure}

\begin{figure}[hbtp]\begin{center}
   \includegraphics{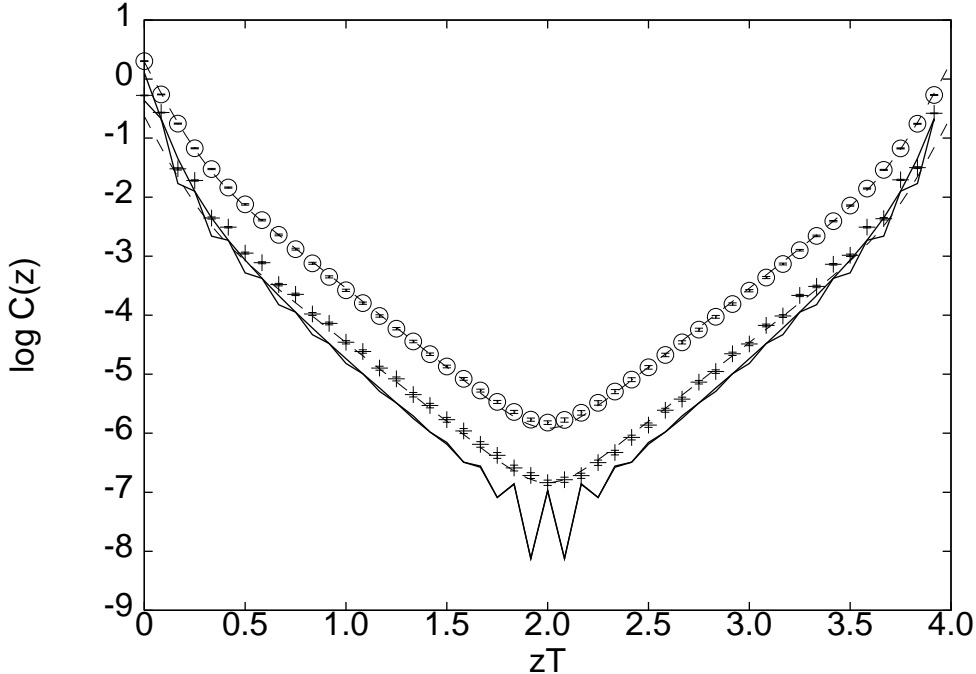}
   \end{center}
   \caption{The $A_1^{++}$ screening correlators in the PS (circle) and
       unprojected V (pluses) channels compared to their ideal gas
       counterparts (heavy lines) for staggered quarks. These were
       determined on $12\times26^2\times48$ lattices at $T=2T_c$ with
       staggered quarks of $m=0.03T_c$. The best fit to the PS data is
       shown (broken line) as is the same fit multiplied by an overall
       constant to superpose it on the data for V, in order to show that
       these two correlators give the same screening mass. Note that the
       short distance part of the V correlator agrees rather well with the
       ideal gas expectation, unlike the PS correlator which differs by a
       factor two.}
\label{fg.corr}\end{figure}

We reiterate the following point. In this quenched work with the Wilson
action we have taken staggered and Naik quarks to provide two different
operators to measure the same physical quantity, namely $\chi_3$. The
agreement between the two provides a check that we have control over the
continuum limit. In this approach we find that equally small lattice
spacings are needed for the two kinds of quarks, and therefore Naik
quarks are fourfold more time consuming.  The use of improved actions to
get continuum results with coarse lattices is a different problem. Our
observations do not rule out the  possibility that improved gauge actions
with fat link dynamical Naik quarks approach the continuum action more
rapidly and that with such actions fat link Naik quark operators have
smoother approach to the continuum already at coarser lattice spacings.

The off-diagonal susceptibility, $\chi_{ud}/T^2$, is consistent with
zero at 95\% confidence level on all the lattice sizes, couplings and
masses investigated, and at the 67\% level on most. The extrapolation to
the continuum of $\chi_{ud}/T^2$ can be performed by the same functional
form as for $\chi_3/T^2$, and, unsurprisingly, gives a result consistent
with zero for both staggered and Naik quarks.  For the strange quark mass
region also, similar results are obtained. In view of these, we neglect
the quark-line disconnected amplitudes in all the susceptibilities in
eq.\ (\ref{chi0q}). The continuum results for $\chi_s/T^2$, $\chi_0/T^2$
and $\chi_Q/T^2$ are discussed in the next section.

We have seen clear degeneracies of the S and PS correlators and the
V and AV correlators--- in agreement with all previous observations.
The new result is that with sufficiently fine lattices spacings, \ie,
$a=1/8T$--$1/12T$, we find a degeneracy of the $A_1^{++}$ masses from the
S/PS correlators and the V/AV correlators. In Figure \ref{fg.coincide}
we demonstrate this by showing that the S/PS correlators are coincident
with the V/AV correlators up to a constant multiplicative factor.

The local masses plateau for distances $z>1/T$ and agree with the results
of the fit. An example is shown in Figure \ref{fg.mpi}. The result of
a two mass fit to the PS correlator coincides within errors with the
plateau in the local masses. Also, the local masses from the parity
projected $A_1^{++}$ components of the V/AV correlators coincide with
those from the S/PS correlators. Our estimates of the common mass are
collected in Table \ref{tb.mass}.

For $a=1/4T$ our extraction of $\mu_{PS}/T$ is completely consistent with
previous estimates; for example, $a\mu_{PS}=1.144$ (4) at $T=2T_c$,
in perfect agreement with the results of \cite{plbold}. However,
with decreasing lattice spacing we find that $\mu_{PS}$ increases
rapidly toward the values expected for an ideal quark gas, eq.\
(\ref{stag}). In Figure \ref{fg.mass} we show $\mu_{PS}/\mu_{FFT}$ as a
function of $1/N_t^2\propto a^2$. Already for $a=1/10T$ the screening
mass is completely consistent with that in an ideal gas. It can be
seen that a quadratic extrapolation to the continuum limit at $2T_c$ is
consistent with an ideal quark gas.  At all three temperatures we see
clear degeneracy of all the $A_1^{++}$ screening masses on the larger
lattices. The $B_1^{++}$ correlators vanish within errors even on the
coarsest lattice \cite{b1pp}.  This behaviour is also compatible with
expectations from an ideal quark gas.

\begin{figure}[htb]\begin{center}
   \includegraphics{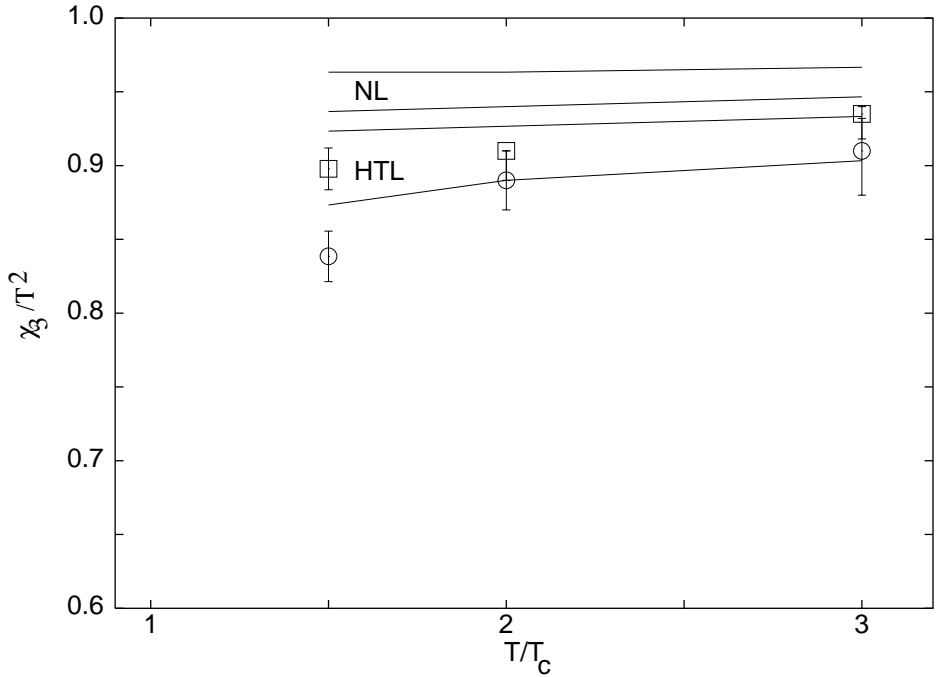}
   \end{center}
   \caption{The ratio $\chi_3/\chi^3_{FFT}$ in quenched QCD, shown as a
       function of $T$. Circles denote results for staggered quarks and
       boxes for Naik quarks. Also shown are the hard thermal loop (HTL)
       and resummed (NL) results from \cite{bir}. The bands for the last
       two are due to uncertainties in $T_c/\lms$.}
\label{fg.tdep}\end{figure}

\begin{table}\begin{center}\begin{tabular}{lrrrrrrr}
   \hline
   $T/T_c$ & $\qquad\chi_3/T^2$ & 
         $\qquad\chi_{ud}/T^2$ & $\qquad\chi_s/T^2$ &
         $\qquad\chi_0/T^2$ & $\qquad\chi_Q/T^2$ &
         $\qquad\lambda_s(T)$ & $\qquad\mu(A_1^{++})/2\pi T$ \\
   \hline
   1.5 & 0.84 (2) & $\quad(-2\pm3)\times10^{-5}$ & 0.53 (1) & 0.43 (1) 
                  & 0.99 (2) & 0.63 (3) & 0.96 (1) \\
   2.0 & 0.89 (2) & $\quad(-4\pm4)\times10^{-6}$ & 0.71 (2) & 0.47 (1) 
                  & 1.07 (2) & 0.80 (3) & 1.01 (1) \\
   3.0 & 0.90 (3) & $\quad(2\pm2)\times10^{-6}$  & 0.84 (3) & 0.49 (1) 
                  & 1.09 (3) & 0.93 (4) & 1.00 (1) \\
   \hline
\end{tabular}\end{center}
\caption{Results for the continuum limit of quark number susceptibilities,
   the Wroblewski parameter, $\lambda_s(T)$, expected of a system at
   thermal equilibrium at temperature $T$, and the $A_1^{++}$ screening
   mass in quenched QCD. We have taken $m_s/T_c=1$, as appropriate to full
   QCD. All the values reported in this table have been obtained by continuum
   extrapolation of our staggered quark results. Those from Naik quarks
   are compatible within 1--$\sigma$ limits, except for $\chi_3$ at
   $T=1.5T_c$ which agrees within 95\% confidence limits.}
\label{tb.res}\end{table}

Interestingly, the measured values of $\langle\overline\psi\psi\rangle$
differ from the ideal gas value by a factor of about two on all the
lattices studied. Within statistical errors $\langle \overline\psi
\psi \rangle$ is independent of volume, and its ratio to the ideal gas
value has small lattice spacing dependence, extrapolating smoothly to
$1.90\pm0.04$ at $2T_c$ in the continuum. In view of the chiral Ward
identity of eq.\ (\ref{ward}) which relates the pion correlator to
the chiral condensate, and the agreement of the screening masses with
their ideal gas values, we examine the screening correlators for the
origin of this major departure from ideal gas behaviour.  In Figure
\ref{fg.corr} are shown the PS and V screening correlators at $T=2T_c$
obtained with lattice spacing $a=1/12T$. It is clear that the correlators
are significantly different from their ideal gas counterparts, not just
in an overall factor (which would be sufficient to explain the chiral
condensate) but also in shape.

\section{Continuum Physics}

Our results for the dependence of the continuum extrapolated values of
$\chi_3/T^2$ on $T/T_c$ are shown in Figure \ref{fg.tdep}. We find that
the results obtained with staggered and Naik quarks are compatible with
each other and with the HTL results of \cite{bir} at the 2--$\sigma$
level. While they are not compatible with the resummed results of
\cite{bir} at the 1--$\sigma$ level, compatibility cannot be ruled out
at the 3--$\sigma$ level for $T\ge2T_c$.  There is a suggestion of a
drop in $\chi_3/T^2$ as one approaches $T_c$. In Table \ref{tb.res}
we have listed the continuum extrapolated values of $\chi_3/T^2$ and
$\chi_{ud}/T^2$ along with the susceptibilities $\chi_s/T^2$, $\chi_0/T^2$
and $\chi_Q/T^2$ which are of relevance to heavy-ion experiments.

Previous lattice investigations at cutoff of $1/4T$ have shown that
the effect of unquenching is to increase $\chi_3/T^2$ by less than 5\%
\cite{qnch,2fl}. The computations of \cite{bir} indicate that the effect
of unquenching is to decrease $\chi_3/T^2$ by less than 5\%. Thus the
effect of including light sea quarks could change the results in Table
\ref{tb.res} either way within a band of 3--5\%.

The Wroblewski parameter \cite{wrobold}, $\lambda_s$, measures the
ratio of primary produced strange to light quark pairs. It has been
argued \cite{wrob} that, for a system undergoing freeze out from
thermal and chemical equilibrium at temperature $T$, $\lambda_s$
is just the ratio $2\chi_s/(\chi_u+\chi_d)$ computed at $T$.  We have
listed the continuum values of $\lambda_s(T)$ obtained from our lattice
simulations in Table \ref{tb.res}. Note especially that $\lambda_s(T)$
for the staggered and Naik quarks are within 1--$\sigma$ of each
other. Various extrapolations of $\chi_u=\chi_d$ and $\chi_s$ yield
$\lambda_s(T_c)\approx0.4$--$0.5$. It seems plausible that at $T\approx
T_c$, $\lambda_s$ dips toward the value obtained from analysis of RHIC
data, \ie, $0.47\pm0.04$ \cite{cleymans}.

Table \ref{tb.res} also contains our determination of the common screening
mass in the $A_1^{++}$ sector for light quarks. At temperatures of $2T_c$
and higher this is completely compatible with a computation in an ideal
gas. Similarly, $\chi_{ud}/T^2$ is also compatible with expectations
from an ideal gas. It has been pointed out before that $\chi_{ud}/T^2$ is
significantly smaller than might be expected in the leading perturbative
treatment of the interactions.  In future it would be interesting to
investigate perturbative corrections to the screening masses and check
whether a similar puzzle exists also for them, or whether interaction
effects indeed are smaller than the error bars on these quantities.
Departures from ideal gas behaviour, by a factor of almost two, has been
observed in the chiral condensate, and hence in the PS susceptibility. It
would be useful to have perturbative predictions for these quantities
in the future.


\begin{thebibliography}{99}
\bibitem{biel}
   G.\ Boyd {\sl et al.\/}, {\sl Nucl.\ Phys.\/}, B 469 (1996) 419.
\bibitem{myold}
   S.\ Gupta, {\sl Phys.\ Rev.\/}, D 64 (2001) 034507.
\bibitem{qcdpax}
   C.\ Bernard {\sl et al.\/}, {\sl Phys.\ Rev.\/}, D 55 (1997) 6861.
   J.\ Engels {\sl et al.\/}, {\sl Phys.\ Lett.\/}, B 396 (1997) 210;
   A.\ Ali Khan {\sl et al.\/}, {\sl Phys.\ Rev.\/}, D 63 (2001) 034502,
     {\sl ibid.\/}, D 64 (2001) 074510.
\bibitem{braaten}
   J.\ O.\ Andersen, E.\ Braaten and M.\ Strickland, {\sl Phys.\ Rev.\/}, D 61 (2000) 014017,
   {\sl ibid.\/}, D 61 (2000) 074016;
   J.\ O.\ Andersen {\sl et al.\/}, hep-ph/0205085.
\bibitem{parwani}
   R.\ Parwani, {\sl Phys.\ Rev.\/}, D 63 (2001) 054014, {\sl ibid.\/}, D64 (2001) 025002.
\bibitem{bir0}
   J.-P.\ Blaizot, E.\ Iancu and A.\ Rebhan, {\sl Phys.\ Rev.\ Lett.\/}, 83 (1999) 2906.
\bibitem{helsinki}
   K.\ Kajantie {\sl et al.\/}, {\sl Phys.\ Rev.\ Lett.\/}, 86 (2001) 10.
\bibitem{saumen}
   S.\ Datta and S.\ Gupta, {\sl Nucl.\ Phys.\/}, B 534 (1998) 392 and hep-ph/9809382;
   S.\ Datta and S.\ Gupta, {\sl Phys.\ Lett.\/}, B 471 (2000) 382.
\bibitem{dimred}
   A.\ Hart, M.\ Laine and O.\ Philipsen, {\sl Nucl.\ Phys.\/}, B 586 (2000) 443.
\bibitem{gluon}
   S.\ Datta and S.\ Gupta, hep-lat/0208001.
\bibitem{qnch}
   R.\ V.\ Gavai and S.\ Gupta, {\sl Phys.\ Rev.\/} D 64 (2001) 074506.
\bibitem{2fl}
   R.\ V.\ Gavai, S.\ Gupta and P.\ Majumdar, {\sl Phys.\ Rev.\/}, D 65 (2002) 054506.
\bibitem{gott}
   S.\ Gottlieb {\sl et al.\/}, {\sl Phys.\ Rev.\ Lett.\/} 59 (1987) 2247.
\bibitem{gavai}
   R.\ V.\ Gavai {\sl et al.\/}, {\sl Phys.\ Rev.\/} D 40 (1989) 2743;
   C.\ Bernard {\sl et al.\/}, {\sl Phys.\ Rev.\/} D 54 (1996) 4585;
   S.\ Gottlieb {\sl et al.\/}, {\sl Phys.\ Rev.\/}, D 55 (1997) 6852.
\bibitem{wrob}
   R.\ V.\ Gavai and S.\ Gupta, {\sl Phys.\ Rev.\/}, D 65 (2002) 094515.
\bibitem{heller}
   C.\ Bernard {\sl et al.\/}, hep-lat/0209079.
\bibitem{koch}
   M.\ Asakawa, U.\ Heinz and B.\ M\"uller, {\sl Phys.\ Rev.\ Lett.\/} 85 (2000) 2072;
   S.\ Jeon and V.\ Koch, {\sl ibid.\/} 85 (2000) 2076;
   D.\ Bower and S.\ Gavin, {\sl Phys.\ Rev.\/} C 64 (2001) 051902;
   S.\ Jeon, V.\ Koch, K.\ Redlich and X.\ N.\ Wang, nucl-th/0105035.
\bibitem{mu}
   Z.\ Fodor and S.\ D.\ Katz, {\sl J.\ High Energy Phys.\/}, 03 (2002) 014;
   C.\ R.\ Allton {\sl et al.\/}, hep-lat/0204010;
   P.\ de Forcrand and O.\ Philipsen, hep-lat/0205016;
   M.\ D'Elia and M.-P.\ Lombardo, hep-lat/0205022.
\bibitem{bir}
   J.-P.\ Blaizot, E.\ Iancu and A.\ Rebhan, {\sl Phys.\ Lett.\/}, B 523 (2001) 143.
\bibitem{4flav} 
  C.\ DeTar and J.\ B.\ Kogut, {\sl Phys.\ Rev.\ Lett.\/}, 59 (1987) 399;
  K.\ D.\ Born {\sl et al.\/}, {\sl Phys.\ Rev.\ Lett.\/}, 67 (1991) 302.
\bibitem{milc} 
  S.\ Gottlieb {\sl et al.\/}, {\sl Phys.\ Rev.\ Lett.\/}, 59 (1987) 1881;
  C.\ Bernard {\sl et al.\/}, {\sl Phys.\ Rev.\/}, D 45 (1992) 3854;
  S.\ Gottlieb {\sl et al.\/}, {\sl Phys.\ Rev.\/}, D 47 (1993) 3619.
\bibitem{gocksch} 
   A.\ Gocksch, P.\ Rossi and U.\ Heller, {\sl Phys.\ Lett.\/}, B 205 (1988) 334.
\bibitem{plbold} 
  S.\ Gupta, {\sl Phys.\ Lett.\/}, B 288 (1992) 171.
\bibitem{boyd} 
  G.\ Boyd, S.\ Gupta and F.\ Karsch, {\sl Nucl.\ Phys.\/}, B 385 (1992) 482;
  G.\ Boyd {\sl et al.\/}, {\sl Z.\ Phys.\/}, C 64 (1994) 331,
  and {\sl Phys.\ Lett.\/}, B 349 (1995) 170.
\bibitem{robert} 
   R.\ V.\ Gavai, S.\ Gupta and R.\ Lacaze, {\sl Phys.\ Rev.\/}, D 65 (2002) 094504.
\bibitem{edwin} 
  T.\ Hashimoto, A.\ Nakamura and I.\ O.\ Stamatescu, {\sl Nucl.\ Phys.\/}, B 406 (1993) 325;
  P.\ de Forcrand {\sl et al.\/}, {\sl Phys.\  Rev.\/}, D 63 (2001) 054501;
  E.\ Laermann and S.\ Schmidt, {\sl Eur.\ Phys.\ J.\/}, C 20 (2001) 541.
\bibitem{old}
  S.\ Gupta, {\sl Phys.\ Rev.\/}, D 60 (1999) 094505.
\bibitem{naik}
   S.\ Naik, {\sl Nucl.\ Phys.\/}, B 316 (1989) 238.
\bibitem{improve}
  K.\ Symanzik, {\sl Nucl.Phys.\/}, B 226 (1983) 187;
\bibitem{chempot}
  P.\ Hasenfratz and F.\ Karsch, {\sl Phys.\ Lett.\/}, B 125 (1983) 308;
  N.\ Bilic and R.\ V.\ Gavai, {\sl Z.\ Phys.\/}, C 23 (1984) 77;
  R.\ V.\ Gavai, {\sl Phys.\ Rev.\/}, D 32 (1985) 519.
\bibitem{b1pp}
  R.\ V.\ Gavai and S.\ Gupta, {\sl Phys.\ Rev.\ Lett.\/}, 83 (1999) 3784.
\bibitem{naikpot}
  R.\ V.\ Gavai, hep-lat/0209008.
\bibitem{wrobold}
  A.\ Wroblewski, {\sl Acta Phys.\ Pol.\/}, B 16 (1985) 379.
\bibitem{cleymans}
  J.\ Cleymans, hep-lat/0201142.
\end{thebibliography}
\end{document}